\documentclass[aip,apl,superscriptaddress,twocolumn,reprint]{revtex4-1}

\usepackage{graphicx}
\usepackage{textcomp}
\usepackage{wasysym}

\begin{document}

\title{Pulse contrast enhancement via non-collinear sum-frequency generation with the signal and idler of an optical parametric amplifier}

\author{E.~Cunningham}
\email{efcunn@slac.stanford.edu}
\author{E.~Galtier}
\author{G.~Dyer}
\author{J.~Robinson}
\author{A.~Fry}
\affiliation{Linac Coherent Light Source, SLAC National Accelerator Laboratory, Menlo Park, CA 94025, USA}

\date{\today}

\begin{abstract}
We outline an approach for improving the temporal contrast of a high-intensity laser system by $>$8 orders of magnitude using non-collinear sum-frequency generation with the signal and idler of an optical parametric amplifier. We demonstrate the effectiveness of this technique by cleaning pulses from a millijoule-level chirped-pulse amplification system to provide $>$10$^{12}$ intensity contrast relative to all pre-pulses and amplified spontaneous emission $>$5~ps prior to the main pulse. The output maintains percent-level energy stability on the time scales of a typical user experiment at our facility, highlighting the method's reliability and operational efficiency. After temporal cleansing, the pulses are stretched in time before seeding two multi-pass, Ti:sapphire-based amplifiers. After re-compression, the 1~J, 40~fs (25~TW) laser pulses maintain a $>$10$^{10}$ intensity contrast $>$30~ps prior to the main pulse. This technique is both energy-scalable and appropriate for preparing seed pulses for a TW- or PW-level chirped-pulse amplification laser system.
\end{abstract}

\maketitle

High-intensity, ultrafast lasers are a valuable tool for the creation of high-energy density (HED) conditions~\cite{Glenzer2016} such as those relevant in fusion research~\cite{Atzeni2004, Lindl2004}, high-energy particle beam production~\cite{Kemp2012}, and astrophysical systems~\cite{Blandford1987, Kugland2012}. To reach the high pulse energies needed for HED experiments, these high-power laser systems employ techniques like chirped pulse amplification (CPA)~\cite{Strickland1985}, optical parametric amplification (OPA)~\cite{Baumgartner1979}, and optical parametric chirped pulse amplification (OPCPA)~\cite{Dubietis1992} to boost the energy of low-power seed pulses by billions of times or more. Unfortunately, this amplification process often introduces undesirable temporal artifacts peripheral to the main pulse, such as pre-pulses, amplified spontaneous emission (ASE), or super-fluorescence. Under the tight focusing conditions required to produce high-intensity laser fields, even low levels of energy arriving early can alter the sample, thus failing to realize the high-energy density condition of interest. Indeed, even a very small level of pre-pulse impinging on a solid target just a few tens of picoseconds ahead of the main pulse can significantly impact the scale of the near-critical plasma density gradient, profoundly affecting the laser-target interaction of the main pulse~\cite{Schollmeier2015}. To avoid such issues, high-intensity laser systems often require temporal contrast enhancement. 

To discriminate against pre-pulses and pulse pedestals, many temporal cleaning techniques exist that rely on nonlinear optical elements whose throughput is much higher for the high-intensity main pulse than for the low-intensity detritus. These nonlinear phenomena include second harmonic generation (SHG)~\cite{Comly1968}, saturable absorption~\cite{Itatani1998, Fourmaux2011}, nonlinear polarization rotation~\cite{Kalashnikov2004}, cross-polarized wave generation (XPW)~\cite{Jullien2005, Antonucci2009, Ramirez2013, Liebetrau2014}, strong-field ionization~\cite{Kapteyn1991, Thaury2007}, and parametric amplification~\cite{Kiriyama2006, Musgrave2010}. These nonlinearities can even be combined -- such as the compounding of SHG with OPA~\cite{Wang1994,Shah2009}, non-collinear OPA with SHG~\cite{Huang2011}, or XPW with SHG~\cite{Yu2018} -- to increase the contrast enhancement.

In this Letter, we present a technique for temporal pulse cleaning based on non-collinear sum-frequency mixing of the signal and idler of an OPA system. This approach transforms the low-contrast output of a high-gain front-end laser system into a high-contrast seed appropriate for further amplification in the back-end of a TW- or PW-class dual CPA system~\cite{Kalashnikov2005}. Beside providing high contrast pulses, this method is straightforward to implement via commercially-available OPA and frequency mixing systems, which allow for reliable, efficient operation.

\begin{figure*}[htb]
  \centering
  \includegraphics[width=6.69in]{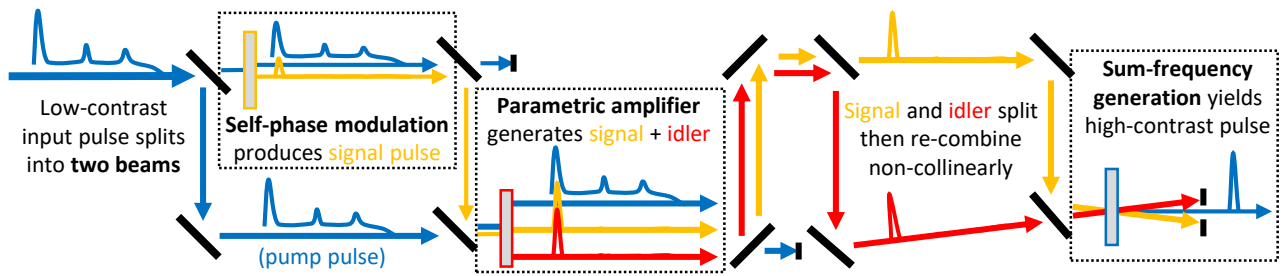}
\caption{Pulse cleaning is achieved through self-phase modulation, parametric amplification, and sum-frequency generation of the generated `signal' and `idler'. The temporal profile of the input pulse (pump), signal, and idler wavelengths are depicted in blue, orange, and red, respectively, at each stage of the system. \label{fig:sketch}}
\end{figure*}

Figure~\ref{fig:sketch} illustrates the general concept of this method. The blue curve in the top-left of the figure represents the temporal profile of a front-end laser system with poor pre-pulse intensity contrast. The pulse is split into two separate beams as part of a white light-seeded OPA system. The weaker of the two split beams is frequency-broadened using \textbf{self-phase modulation} (SPM; shown by the orange curve in the first box) to produce a long-wavelength tail in the spectrum. This portion of the broadened spectrum serves as the `signal' in an ensuing \textbf{parametric amplifier}, which is depicted in the second box. Here, the stronger of the two split beams acts as the `pump' (blue curve) to generate a high-energy signal (orange curve) and `idler' (red curve) pair through difference-frequency generation (DFG). After discarding the residual pump beam with a spectral filter, the signal and idler of the OPA system are split and then re-combined non-collinearly in a mixing crystal allowing \textbf{sum-frequency generation} (SFG; depicted in the third box). This produces a new, temporally-cleaned pulse -- propagating between the signal and idler beams -- with the same wavelength as the original beam.

The contrast enhancement derives from each of the three optical nonlinearities implemented:

\begin{enumerate}
\item \textbf{Self-phase modulation}: In producing the long wavelength of the signal, the output of the main pulse exceeds that of the pre-pulses in two ways: i) the frequency shift due to SPM is proportional to intensity, and ii) the frequency shift due to SPM is also proportional to the propagation length, which can be longer for the high-intensity main pulse because of self focusing and filamentation~\cite{Boyd1992a}.
\item \textbf{Parametric amplification}: In producing the high-energy signal and idler pair, the output of the main pulse exceeds that of the pre-pulses in two additional ways: i) the intensity output of an OPA system is proportional to the input intensity of the signal, and ii) the intensity output of an OPA system depends exponentially on the input intensity of the pump~\cite{Boyd1992b}.
\item \textbf{Non-collinear sum-frequency generation}: In producing the high-contrast pulse with the same wavelength as the input beam, the output of the main pulse exceeds that of the pre-pulses in that the intensity output of the SFG is proportional to the product of the input intensities of the signal and the idler~\cite{Boyd1992c}. In addition, the non-collinear angle of the SFG allows an aperture to block the signal and idler beams {\em along with} any unwanted, collinear components at the same wavelength as the main SFG pulse -- such as super-fluorescence, leakage of the pump beam through the spectral filters, or unintentional SHG of the signal or idler in the mixing crystal -- that would otherwise mar the intensity contrast. In this way, the non-collinear SFG is not only meant to provide a final step of contrast enhancement, but it is also intended to prevent any undoing of the more-significant temporal cleaning already accomplished through SPM and OPA.
\end{enumerate}

The OPA+SFG method possesses several advantages relative to other pulse cleaning approaches. 

\begin{itemize}
    \item By implementing SFG of the signal {\em and} idler outputs of an OPA system instead of using SHG of just one beam, this method utilizes the full output energy of the OPA system, thus improving the overall conversion efficiency. The dual-beam output also allows simpler implementation of the non-collinear frequency mixing geometry, which avoids reliance on spectral filtering for contrast enhancement. Finally, using both OPA beams grants the freedom to tune the signal and idler wavelengths for maximum energy efficiency since the signal and idler sum together to yield the pump wavelength {\em by definition}, whereas a single beam must be tuned specifically to twice the pump wavelength in order to recover the original input wavelength after SHG.
    \item Unlike XPW, this method does not require any optics for separating the final beam, and therefore its ultimate contrast performance is not limited by imperfect polarizers and other difficulties inherent to the extraction of a small signal from a large residual background of the same wavelength. Additionally, this method adopts second-order nonlinearities for its high-energy interactions rather than third-order nonlinearities, which tend to have greater difficulties with optical damage, energy scaling, unwanted additional nonlinearities, and instabilities from intensity fluctuations.
    \item Compared to techniques employing SHG before OPA, the OPA+SFG method benefits from the powerful pre-pulse discrimination provided by the white-light generation stage which produces the new `signal’ seed through SPM. Additionally, the OPA+SFG approach can leverage a stabilizing effect on the output energy by saturating the optical parametric amplifier, whereas some SHG+OPA techniques specifically prescribe low gain in the OPA system. Finally, the OPA+SFG method involves mixing a signal and idler that have {\em both} been temporally cleaned already through SPM and DFG; on the other hand, some SHG+OPA techniques provide a pump pulse cleaned only using simple SHG while the signal pulse is not cleaned at all, and this limits the contrast cleaning achievable in the output idler pulse~\cite{Shah2009}.
    \item This method is comparatively easy to implement even in laboratories without extensive laser expertise due to the commercial availability of well-engineered systems that support \textmu J- and mJ-level OPA and SFG for common short-pulse laser wavelengths (such as 800~nm, 1030~nm, and 1064~nm). 
\end{itemize}

\begin{figure}[h!]
  \centering
  \includegraphics[width=2.5in]{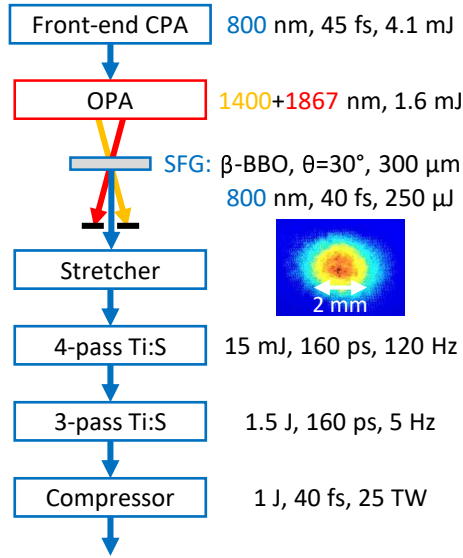}
\caption{Block diagram of the OPA+SFG setup and the back-end TW-class CPA system.\label{fig:block}}
\end{figure}

The OPA+SFG contrast cleaning technique was enacted using the short-pulse laser system in the Matter in Extreme Conditions (MEC) end-station of the Linac Coherent Light Source (LCLS) at SLAC National Accelerator Laboratory~\cite{Nagler2015}. A block diagram of the process is outlined in Fig.~\ref{fig:block}. The output of a commercial Ti:sapphire-based oscillator (Coherent Vitara) seeds a commercial Ti:sapphire-based CPA system (Coherent Legend), which produces 45~fs, 4.1~mJ pulses at a repetition rate of 120~Hz. These pulses seed a commercial OPA system (Light Conversion TOPAS-Prime), which generates white light and then amplifies the long signal wavelength in two DFG crystals. While the energy of the combined signal and idler remains largely constant over the full wavelength tuning range, the maximum output of $\sim$0.9~mJ and $\sim$0.7~mJ centers near 1400~nm and 1866~nm for the signal and idler, respectively. After filtering out the residual pump using dichroic mirrors, the collimated, collinear, orthogonally-polarized signal and idler beams are injected into a commercial frequency-mixing accessory (Light Conversion NDFG). Here, the two beams are separated and then recombined inside a $\beta$-BBO crystal ($\theta=30^\circ$, 300-\textmu m thick) phase matched for type-II SFG. The recombination angle is controlled to insure distinct separation between the residual signal/idler beams and the SFG beam before entering the back-end CPA system, whose stretcher and amplifiers are several meters away. Without re-sizing the beams for higher intensity, the SFG process from a collimated, $\sim$2~mm beam still yields an energy of over 250~\textmu J, translating to an overall efficiency of $\sim$6\%. (This efficiency has been observed as high as $\sim$9\%, with an output of 370~\textmu J, using a loose focus that remains far below the damage threshold of the BBO.) The pulse duration is measured to be 40~fs using a commercial single-shot autocorrelator (SSA).

\begin{figure}[h!]
  \centering
  \includegraphics[width=3.37in]{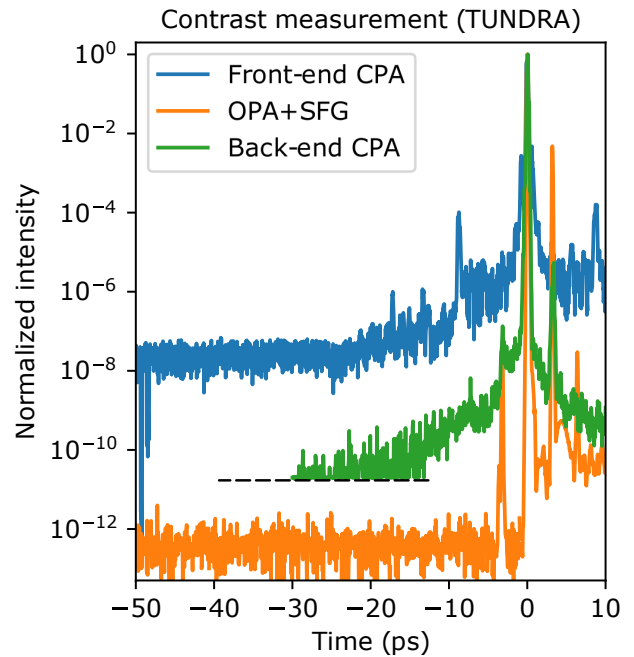}
\caption{Contrast comparison of the front-end Ti:sapphire CPA system before (blue) and after (orange) pulse cleaning. The contrast of the 25~TW beam is shown in green; the dashed line represents the noise floor of this measurement.\label{fig:contrast}}
\end{figure}

The pulse contrast of the output of the front-end CPA system is compared to the output after cleaning with the OPA+SFG setup, as measured by a commercial third-order scanning autocorrelator (Ultrafast Innovations TUNDRA). Figure~\ref{fig:contrast} displays the improvement at the picosecond time scale, with the blue and orange curves plotting the contrast before and after cleaning, respectively. The overall contrast is improved at least to $>$10$^{12}$, as limited by the noise floor of the detector, for all times $>$5~ps before the main pulse. The relative magnitude of the pre-pulse at $t=-10$~ps is shown to decrease by more than eight orders of magnitude, which sets a lower bound on the effectiveness of this technique. The pre-pulse seen at $t=-3$~ps is a possible artifact of the detector, as the same peak has been measured with this device on an independent OPCPA laser system, but this has not been possible to conclusively verify. On longer timescales, inspection of the OPA+SFG output using a bare photodiode and oscilloscope exhibited no traces of ASE or pre-pulses remaining from the front-end CPA system over a detection depth of at least six orders of magnitude.

\begin{figure}[h!]
  \centering
  \includegraphics[width=3.37in]{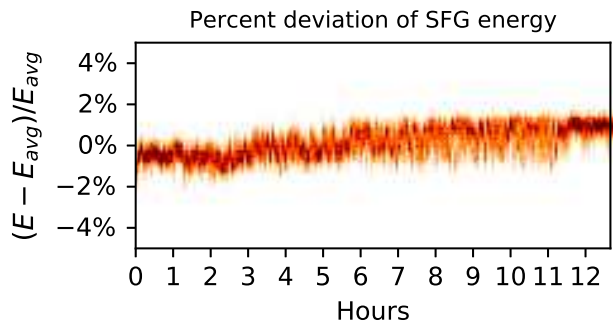}
\caption{Percent deviation of the OPA+SFG energy of 547,000 shots logged at 12~Hz. The overall stability is 0.5\% RMS. The stability of the front-end CPA system during this measurement was 0.2\% RMS. \label{fig:sfg_en}}
\end{figure}

Figure~\ref{fig:sfg_en} characterizes the OPA+SFG output in terms of its energy stability -- a profoundly important parameter to a user facility like the LCLS, where instrument reliability and machine up-time are essential to completing experiments during their brief allocated beam-times. In contrast with the inconsistent performance and near-constant attention required using other pulse cleaning techniques implemented previously, the OPA+SFG setup at MEC maintains an output energy stability on the order of 1\% over the course of twelve hours, which corresponds to the shift length of an LCLS beam-time. Over the course of an entire experimental run (usually several weeks to several months long), performance of the OPA+SFG can be preserved with infrequent, sub-millimeter adjustments to motorized stages controlling the length of the compressor in the front-end CPA system and the temporal delays within the OPA system, neither of which significantly affect the alignment downstream. Furthermore, operation of the OPA+SFG setup has been recoverable immediately even after months of disuse due to extended facility maintenance periods or experimental campaigns using other laser systems. Overall, the excellent stability and operational efficiency of the OPA+SFG setup render it to be ideally suitable for the intense, on-demand requirements of a user facility.

\begin{figure}[h!]
  \centering
  \includegraphics[width=3.37in]{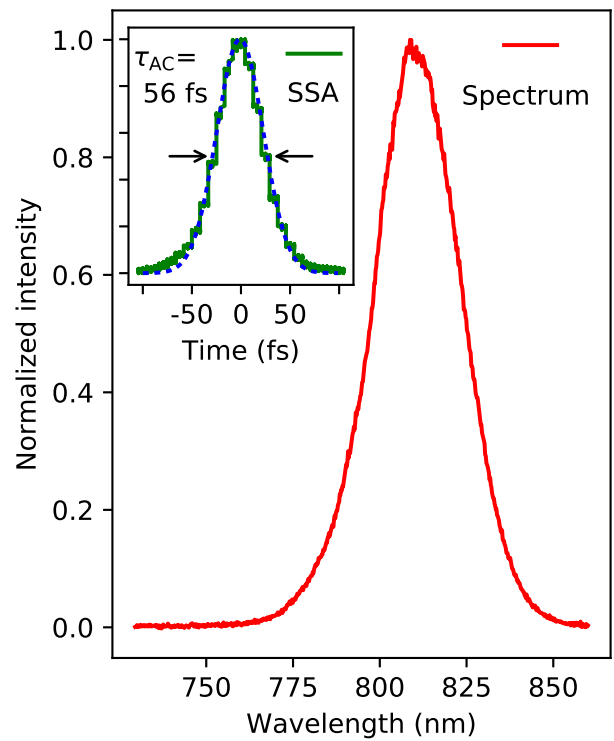}
\caption{Spectrum (red) and single-shot autocorrelation (inset, green) traces measured after compression to 25~TW. \label{fig:output}}
\end{figure}

Demonstrating the suitability of these high-contrast pulses for seeding TW- or PW-class laser systems, the output of the OPA+SFG seeds a back-end, Ti:sapphire-based CPA system in the MEC end-station. First, the pulse is stretched to $\sim$160~ps in a Martinez-type stretcher~\cite{Martinez1987} with a 1500~l/mm grating set at $\sim$56$^\circ$. The seed is then amplified in two multi-pass amplifiers (MPAs). The first MPA yields 15~mJ pulses after four passes through an antireflection-coated $\diameter 8~\mathrm{mm}\times15$~mm-long rod pumped on each side by $\sim$40~mJ from individual diode-pumped, frequency-doubled Nd:YLF systems (Coherent Evolution-HE) operating at 120~Hz. The second MPA produces $>$1.5~J pulses after three passes through an antireflection-coated $\diameter 30~\mathrm{mm} \times 20$~mm-long rod pumped on each side by $\sim$3~J from the beamsplit-output of a flashlamp-pumped, frequency-doubled Nd:YAG system (Thales Gaia) operating at 5~Hz. Both crystals are housed in water-cooled mounts in which the chilled liquid circulates in direct contact with the edges of the amplifier rod. The pulse is then compressed under vacuum using a Treacy compressor~\cite{Treacy1969} with two 1480~l/mm gratings set at $\sim$55$^\circ$. The output pulse features an energy of over 1~J in 40~fs, corresponding to a peak power of at least 25~TW. The spectrum and SSA trace are displayed in Fig.~\ref{fig:output}. The intensity contrast is still measured to be $>$10$^{-10}$ at $t=30$~ps, as seen in Fig.~\ref{fig:contrast}. While outside the scope of this paper, further investigation into the contrast degradation in the back-end CPA system is appropriate, with efforts underway in characterizing the effects of B-integral~\cite{Didenko2008} and scattering~\cite{Li2017} in this particular system.

In conclusion, we have demonstrated a temporal pulse cleaning method that suppresses pre-pulses and other undesirable temporal artifacts by at least eight orders of magnitude to $<$10$^{-12}$ of the main pulse. The technique is operationally robust while also easy to implement with multi-cycle pulses at the mJ level using commercial systems available for several common central wavelengths. As such, the OPA+SFG technique represents an ideal solution for providing a high-contrast seed suitable for amplification of ultrashort pulses to TW- or PW-level peak powers. In some cases, this method should be scalable to higher energies more easily than other temporal cleaning techniques (e.g. XPW) because the relatively-low intensities needed for the second-order nonlinearity are easier to manage, and large-size OPA and SFG crystals are relatively common for several of the wavelengths of most widespread interest. The MEC instrument is supported by the U.S. Department of Energy, Office of Science, Office of Fusion Energy Sciences under Contract No. SF00515.

\end{document}